\documentclass[aps,prl,twocolumn,superscriptaddress,showpac]{revtex4-1}
\usepackage{graphicx}
\usepackage{amsmath}
\usepackage{subfigure}
\usepackage{color}

\begin{document}
\title{The onset of tree-like patterns in negative streamers}

\author{M. Array\'as} \affiliation{\'Area de Electromagnetismo, Universidad Rey Juan Carlos, Camino del Molino s/n, 28943 Fuenlabrada, Madrid, Spain}
%\email{manuel.arrayas@urjc.es}
\author{M. A. Fontelos}
%\email{marco.fontelos@icmat.es}
\affiliation{Instituto de Ciencias Matem\'{a}ticas
(CSIC-UAM-UCM-UC3M), C/ Nicol\'{a}s Cabrera, 28049 Madrid, Spain} %
\author{U. Kindel\'an}
%\email{ultano.kindelan@upm.es}
\affiliation{Departamento de Matem\'{a}tica Aplicada y Met.~Inf, Universidad Polit\'{e}cnica de Madrid, Alenza 4, 28003 Madrid, Spain}

\begin{abstract}
We present the first analytical and numerical studies of the initial
stage of the branching process based on an interface dynamics streamer
model in the fully 3-D case. This model follows from fundamental
considerations on charge production by impact ionization and balance
laws, and leads to an equation for the evolution of the interface
between ionized and non-ionized regions. We compare some experimental
patterns with the numerically simulated ones, and give an explicit
expression for the growth rate of harmonic modes associated with the
perturbation of a symmetrically expanding discharge. By means of full
numerical simulation, the splitting and formation of characteristic
tree-like patterns of electric discharges is observed and
described.
\end{abstract}

\date{\today }
\maketitle

\begin{figure}
\centering
\includegraphics[width=0.2\textwidth]{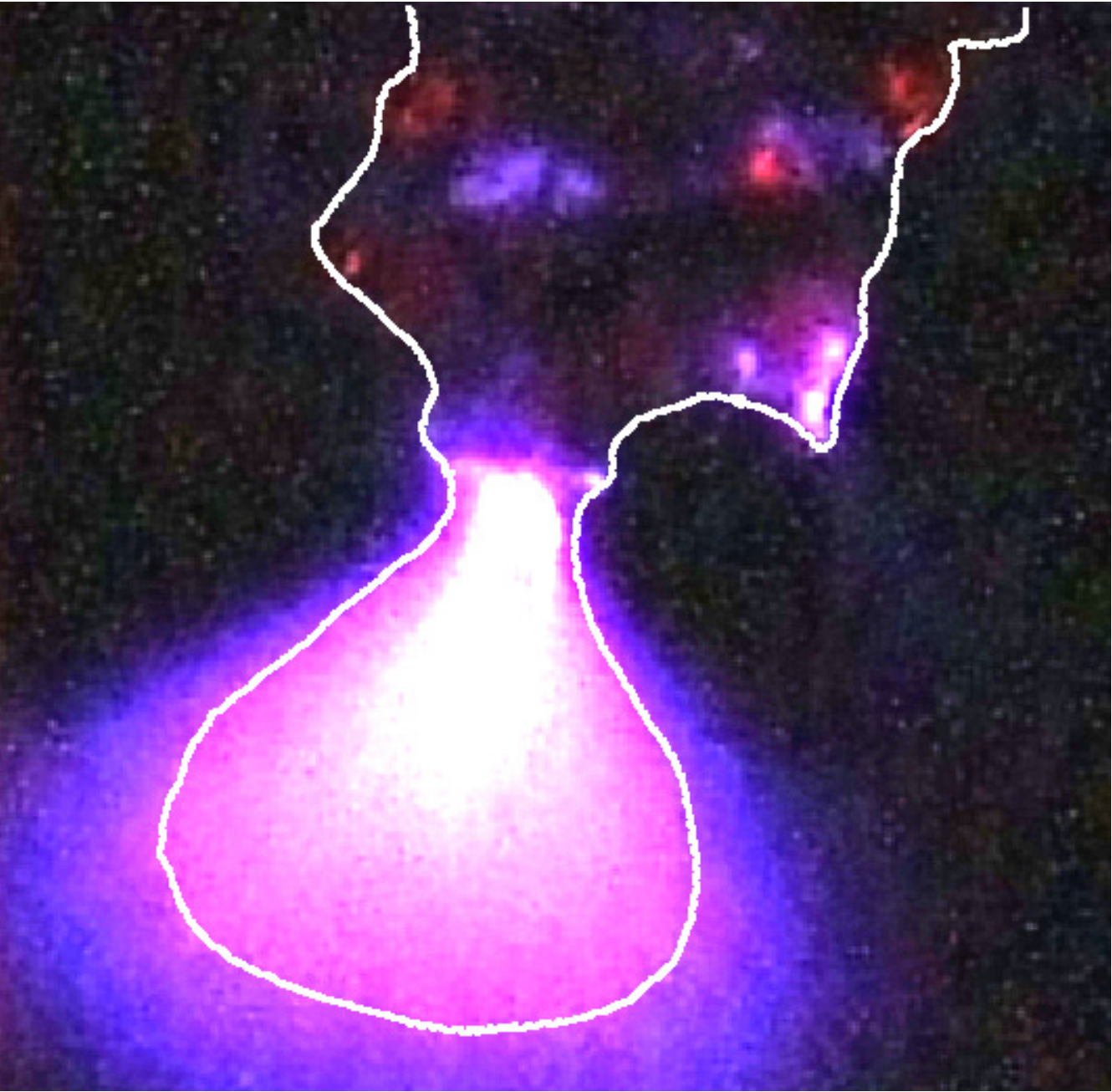}\hspace{0.15cm}\includegraphics[
width=0.2\textwidth]{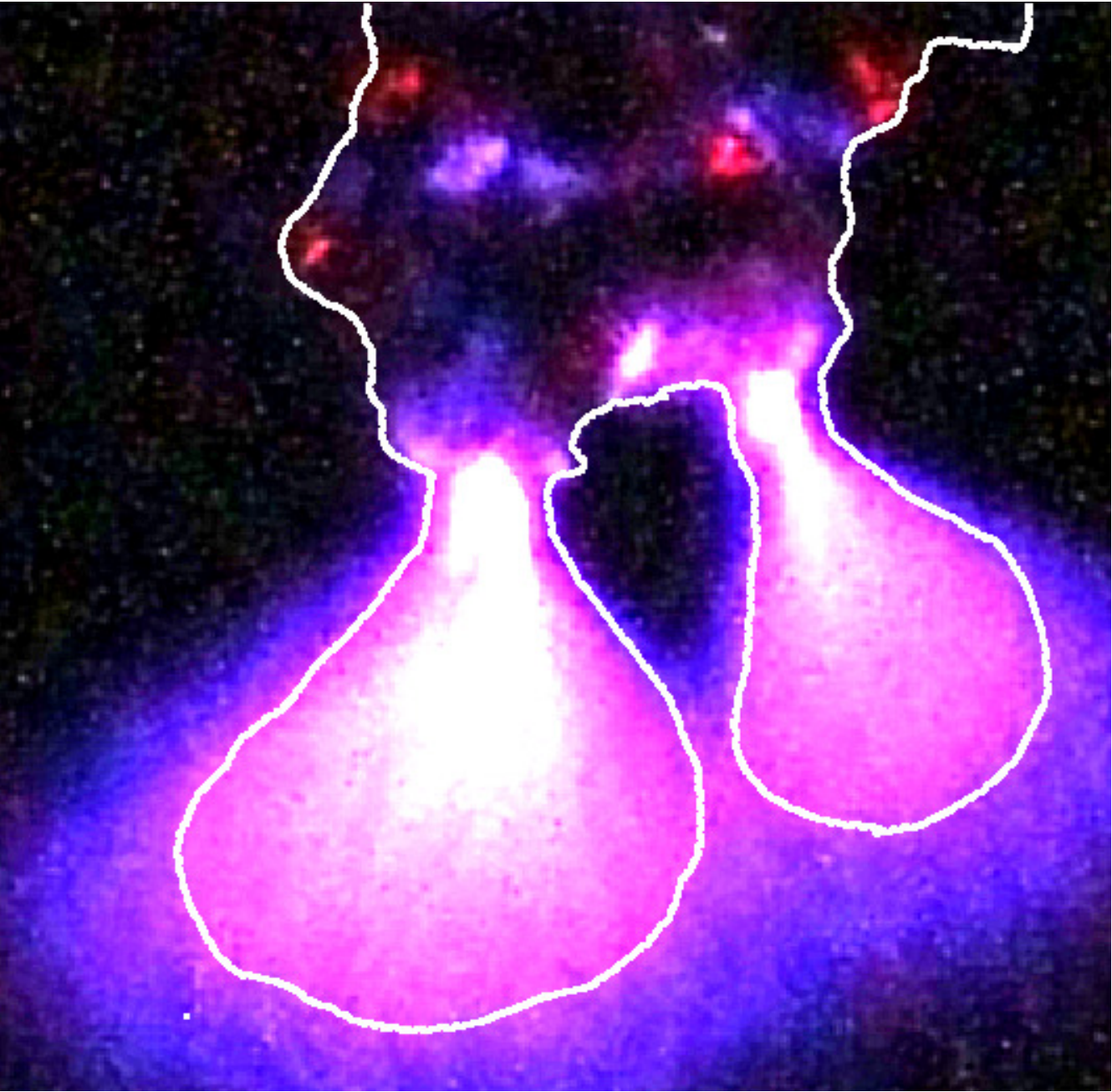}
\caption{Experimental sequential images of the growth of a streamer in
a corona discharge at its early stages. This is a 100 kV point-plane discharge
in the vertical direction with an air-gap of 30 cm. The contour has been highlighted as a visual aid.}
\label{descarga}
\end{figure}

It is a well known visible fact that electric discharges form
tree-like patterns, very much as those in coral reefs and
snowflakes. The study of the branching process leading to such pattern
is of considerable interest both from pure and applied points of view.
Many industrial techniques, ranging from lasers to chemical processing
of gases and water purification could be improved provided the
development of tree-like patterns can be controlled or avoided.
Although an electric discharge is a very complex phenomenon, with
radiation and chemistry processes involved \cite{Rai,Raether,Pasko}, the
description of its initial stage is simpler. A single free electron
traveling in a strong, uniform electric field ionizes the gaseous
molecules around it, generating more electrons and starting a chain
reaction of ionization. The ionized gas creates its own electric
field, which speeds up the reaction, and a streamer is born. The
streamers of ionized gas have an inevitable tendency to break up at
their tips (see figure \ref{descarga}), followed by the creation of the
familiar tree-like pattern .

Early efforts \cite{Dhali, Vitello,Lagarkov, vanS} were able to
identify a minimal streamer model with which, after numerical
simulations under the hypothesis of cylindrical symmetry, an
instability was observed \cite{ME}. Later on, the dispersion relation
for planar fronts was computed and the existence of an instability
leading to the development of fingers was found \cite{aftprl}. Due to
the enormous difficulty for performing full numerical simulations of
the minimal streamer model, some simplified descriptions have been
attempted in the last years (see \cite{ute2} for a review where various ad hoc assumptions are discussed). In any case, the
fully 3-D case has resisted the attack so far. As an alternative
approach (the one we follow in this work), the motion and propagation
of the streamer discharge has recently been described by a contour
dynamics model first introduced in \cite{Arrayas10} and used to
predict with success some experimental features of discharges on
dielectric surfaces \cite{Arrayas11, Japos}. The contour dynamics model
describes the interface separating a plasma region from a neutral gas
region. For a negative discharge, the separating surface has a net
charge $\sigma $ and the thickness goes to zero as $\sqrt{D}$, with
$D$ the charge diffusion coefficient. The interface moves with a
velocity in the normal direction
\begin{equation}
v_{N}=-\mu _{e}\mbox{E}_{\nu }^{+}+2\sqrt{\frac{D_{e}}{l_{0}}\mu _{e}|%
\mbox{E}_{\nu }^{+}|\exp \left( -\frac{\mbox{E}_{ion}}{|\mbox{E}_{\nu }^{+}|}%
\right) }-D_{e}\kappa ,  \label{cm1}
\end{equation}%
where $\mbox{E}_{\nu }^{+}$ is the normal component of the electric field at
the interface when approaching it from outside the plasma region, $\mu _{e}$
the electron mobility, $D_{e}$ is the electron diffusion coefficient, E$_{ion}$
is a characteristic ionization electric field and $\kappa $ is twice the
mean curvature of the interface. The parameter $l_{0}$ is the microscopic
ionization characteristic length.

The total negative surface charge density at the interface changes according
to
\begin{equation}
\frac{\partial \sigma _{e}}{\partial t}+\kappa v_{N}\sigma _{e}=-\frac{%
\mbox{E}_{\nu }^{-}}{\varrho _{e}}-j_{\nu }^{-}\,,  \label{cm2}
\end{equation}%
where $\mbox{E}_{\nu }^{-}$ is the electric field at the interface coming
from inside the plasma, $\varrho _{e}$ is a parameter proportional to the
resistivity of the electrons in the created plasma and $j_{\nu }^{-}$ is the
current contribution at the surface of any source inside the plasma. For
instance, an insulated wire inside the plasma at $\mathbf{x}_{0}$, carrying
an
electric current $I(t)$, will create a current density inside the plasma and
as quasineutrality is fulfilled, we will have for the plasma region
\begin{equation}
\nabla \cdot \mathbf{j}=I(t)\delta (\mathbf{x}-\mathbf{x}_{0})
\label{nablaj}
\end{equation}%
and $\mathbf{j}$ is obtained solving that equation. Note that at the
interface there is an electric field discontinuity given by
\begin{equation}
\mbox{E}_{\nu }^{+}-\mbox{E}_{\nu }^{-}=-\frac{e\sigma }{\varepsilon _{0}}.
\label{jump}
\end{equation}%
%We could use in principle the same model for a positive front, but the
%electric field should be sign reversed, and $\sigma $ would represent the
%positive surface charge density. At the end of this Letter we will discuss
%the case of positive streamers and their qualitative differences with respect
%to negative streamers that have been observed in experiments.

%(although the moving carriers in the model
%are the electrons, one may think of a front made of \textit{holes}) and
%characterized with the corresponding parameters for the mobility, diffusion
%and so on.

\begin{figure}[htbp]
\begin{center}
\includegraphics[scale=0.3,angle=180]{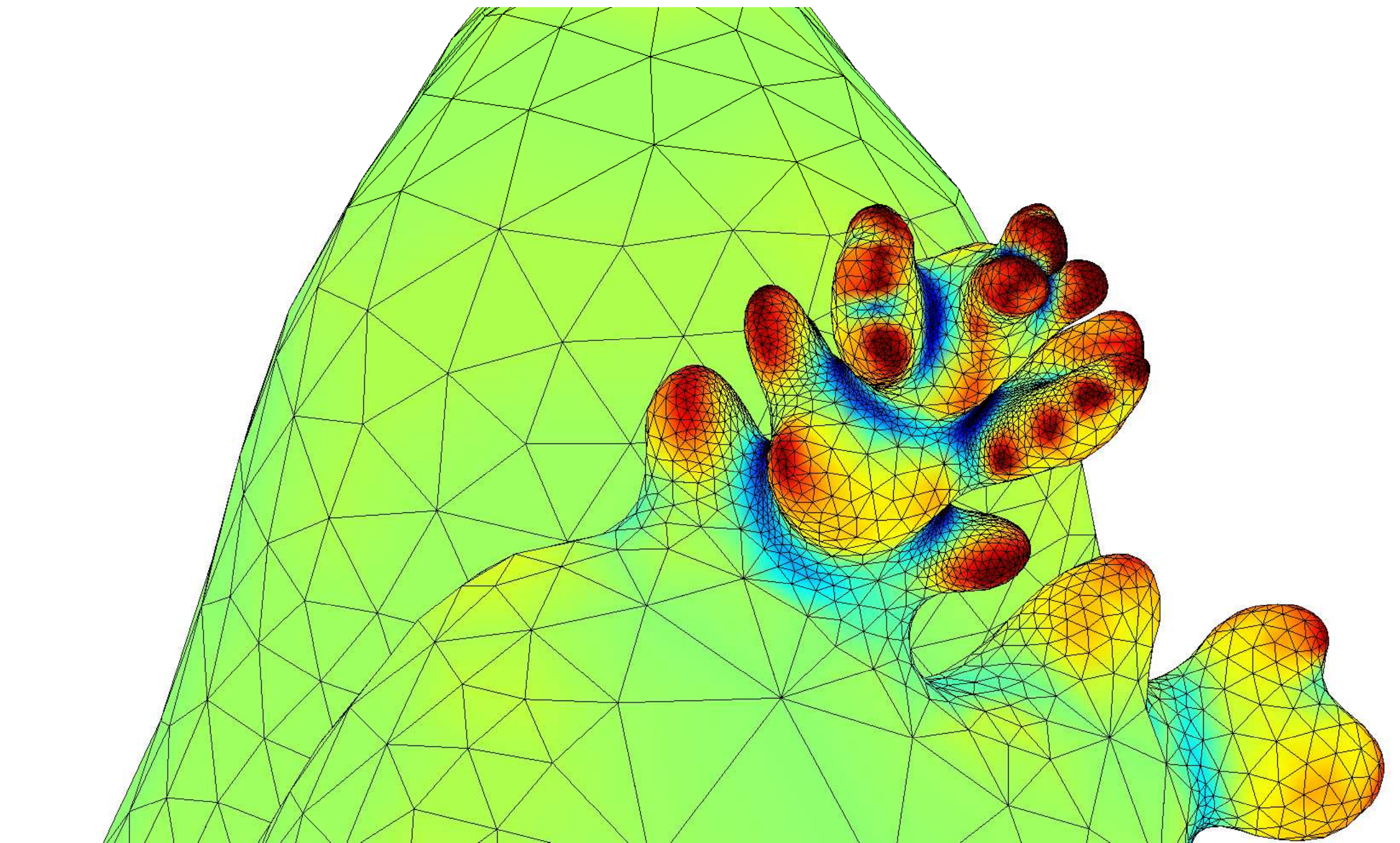}
\end{center}
\caption{Detail of the shape of the plasma at $t=0.1769$}
\label{zoom}
\end{figure}

It is convenient to express the model in dimensionless units. The physical
scales are given by the ionization length $l_0$, the characteristic impact
ionization field $\mbox{E}_{i}$, and the electron mobility $\mu_e$. The
velocity scale yields $U_0=\mu_e\mbox{E}_{i}$, and the time scale $%
\tau_0=l_0/U_0$. Typical values of these quantities for nitrogen at normal
conditions are $l_0\approx 2.3\,\mu\mathrm{m}$, $\mbox{E}_{i} \approx 200$
kV/m, and $\mu_e \approx 380\,\mathrm{cm^2/Vs}$. The unit for the negative
surface density reads $\sigma_0=\varepsilon_0 {\mbox E}_i/e$, so for the
current density $j_0= \sigma_0 U_0/l_0$ and for the resistivity $%
\varrho_0=\mu_el_0/\sigma_0$. The diffusion constant unit turns out $D_0=l_0
U_0$. Introducing dimensionless units, the model reads
\begin{equation}
v_{N}=-\mbox{E}_{\nu }^{+}+2\sqrt{\varepsilon\alpha(|\mbox{E}_{\nu }^{+}|)}%
-\varepsilon\kappa,  \label{bb1}
\end{equation}

\begin{equation}
\frac{\partial \sigma }{\partial t}+\kappa v_{N}\sigma =-\frac{\mbox{E}_{\nu
}^{-}}{\varrho}-j_\nu^{-}\,,  \label{sigmaeq}
\end{equation}%
being
\begin{equation}
\alpha(|\mbox{E}_{\nu }^+|)=|\mbox{E}_{\nu }^+|\exp\left(-\frac{1}{|\mbox{E}%
_{\nu }^+|}\right),  \label{ft}
\end{equation}
and $\varepsilon=D_e/D_0$ the dimensionless diffusion coefficient. In what
follows all the quantities are dimensionless unless otherwise indicated. We
have used an adaptive boundary element method, developed for general contour
dynamics problems (\cite{font1,font2}) in order to perform numerical
simulations with equations \eqref{bb1} and \eqref{sigmaeq}.
\begin{figure}[htbp]
\begin{center}
\subfigure[Initial configuration]{\label{p1gauss}\includegraphics[scale=0.25,angle=180]{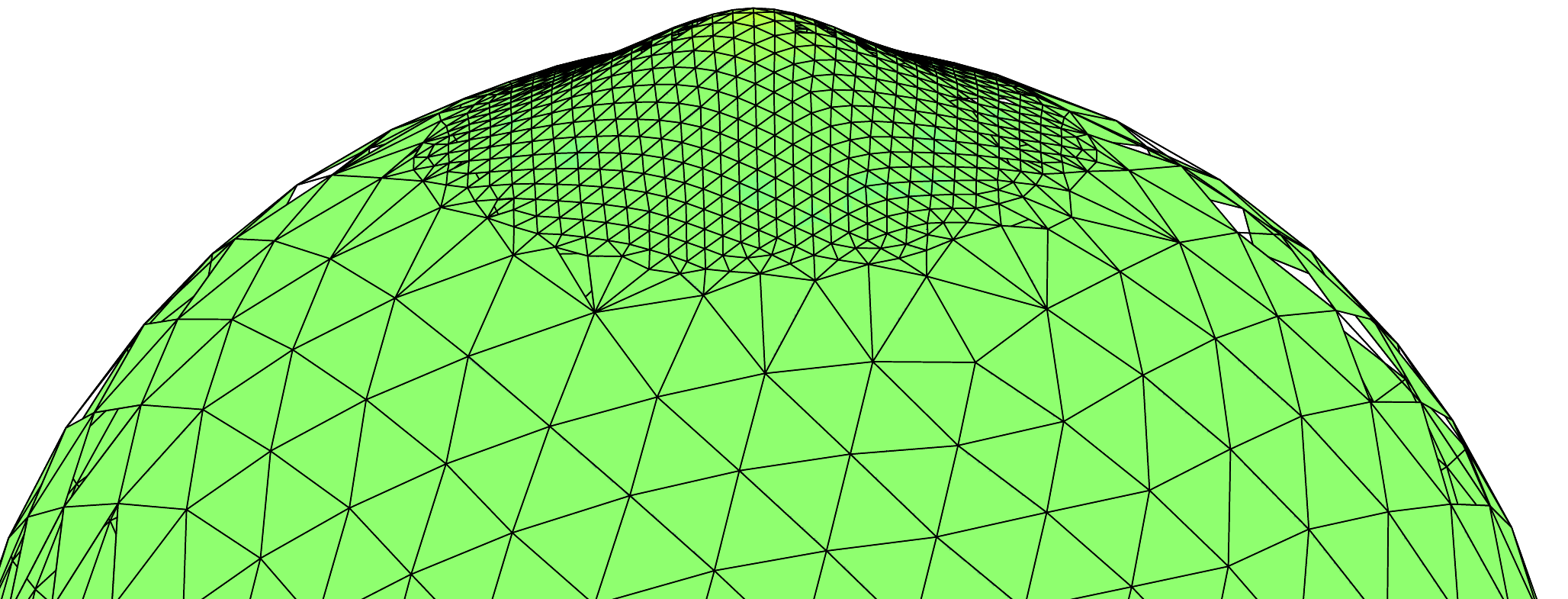}}\\[0.1cm]
\subfigure[$t=0.0794$]{\label{eq25e05_150}\includegraphics[scale=0.25,angle=180]{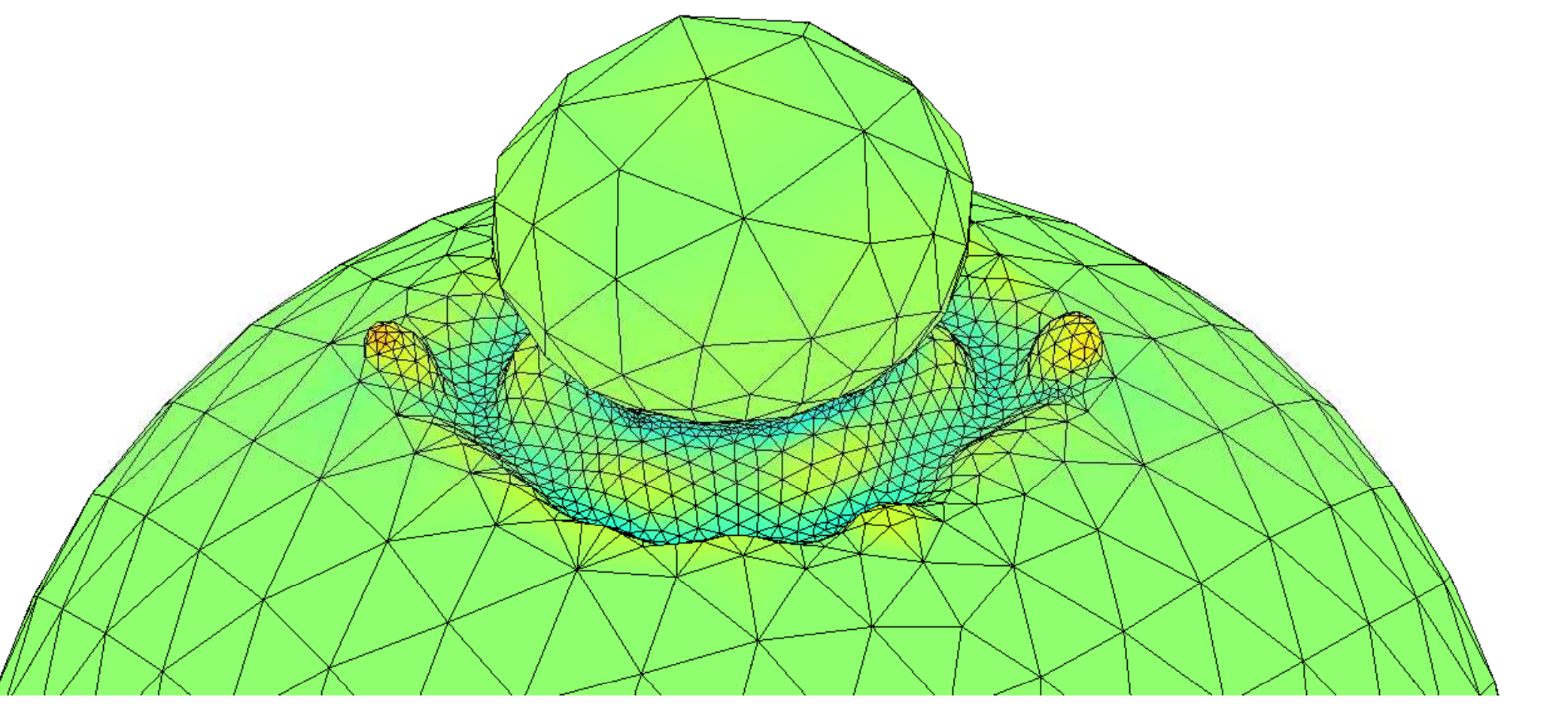}}\\[0.1cm]
\subfigure[$t=0.1438$]{\label{eq25e05_350}\includegraphics[scale=0.25,angle=180]{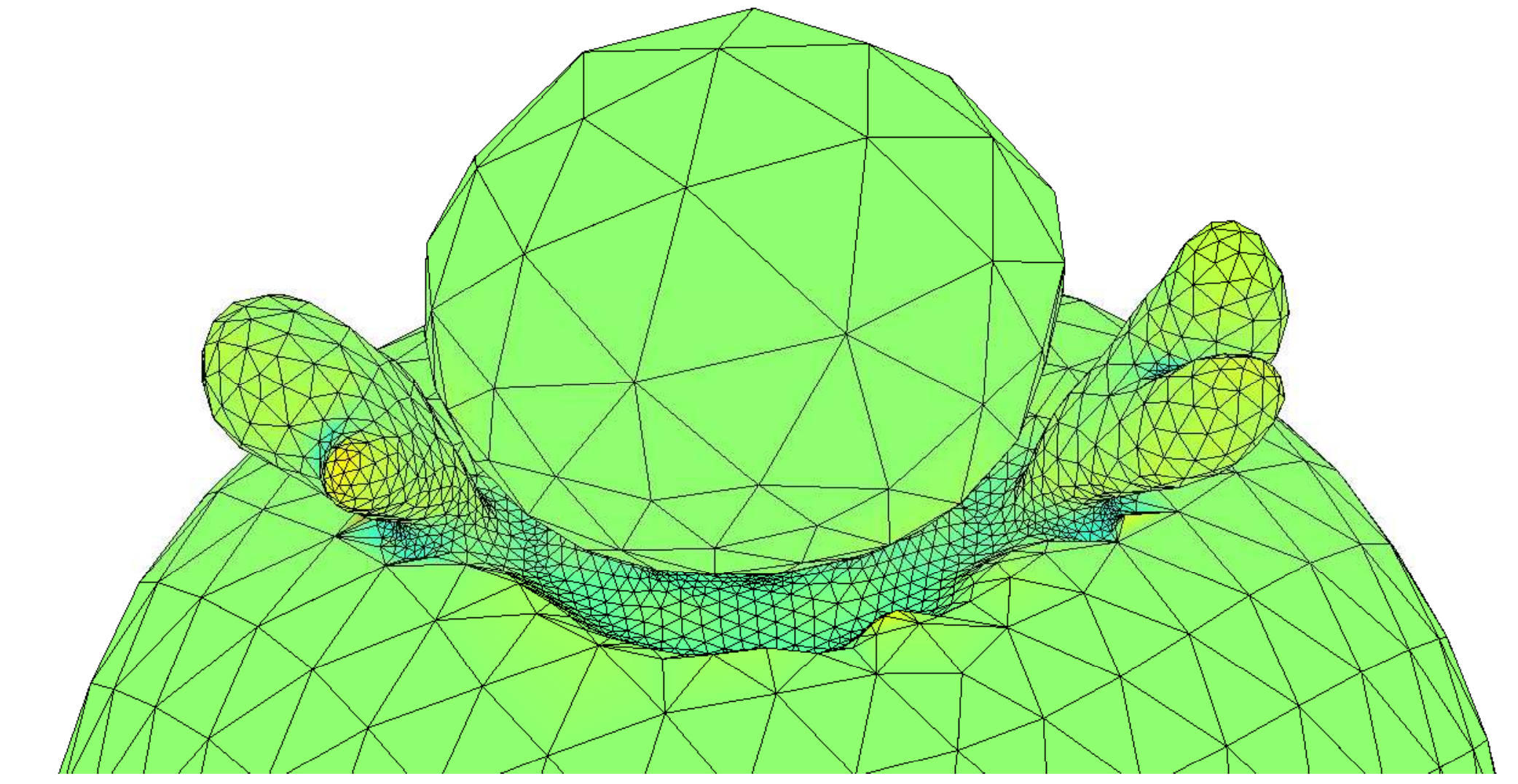}}\\[0.1cm]
\subfigure[$t=0.1769$. The encircled region is enlarged in figure
\ref{zoom}]{\label{eq25e05_1590}
\includegraphics[scale=0.3,angle=180]{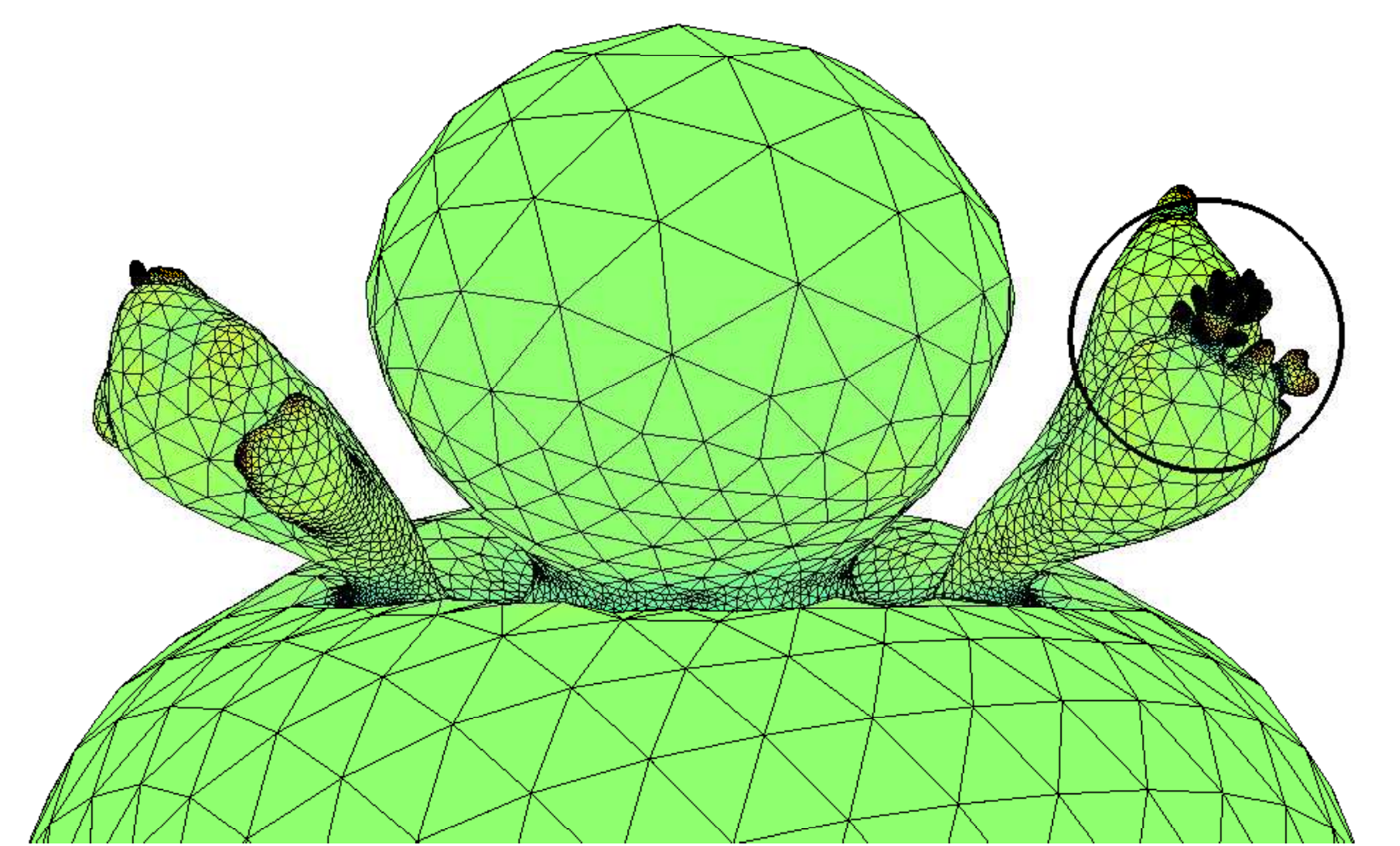}}
\end{center}
\caption{Evolution of the plasma for $Q=-25$, $E=0.5$ and $\varepsilon = 0.02$.
Color gradation represents curvature.}
\label{q25_e05_evol}
\end{figure}
In figure \ref{q25_e05_evol} we show numerical simulations of the evolution of the
discharge at four time steps. The plasma is assumed ideally conducting,
initially charged with integrated surface charge Q=-25, subject
to an external field in the vertical direction $E=0.5$, and confined inside an
initially spherical
geometry perturbed by
$r_0(\theta,
\phi)=R_0+\delta_0(\exp(-(\cos^2(\phi)+\cos^2(\theta))/c)$, with
$c=0.03$ and $\delta_0=0.1$. We first observe the onset of streamer fingers. At
time t = 0.17 the streamers develop further instabilities and split again.
Qualitatively the process can be described in the following terms: any
protuberance that develops is accompanied by an increase of the charge
density at its tip. The electrostatic repulsion of charges at the tip tends
to make the tip expand and the finger grow. In opposition to this is the
action of the surface tension tending to flatten the protuberance and
setting up a flux of charge from the protuberance out to the sides. However,
overall, the protuberance becomes amplified. This process occurs again and
again until a tree-like pattern is produced. In figure \ref{zoom} we depict a
detail of this pattern. Those ideas where anticipated
in \cite{ME}, but due to the restriction of 2-D simulations the whole
process of the branching pattern formation could not be observed.
In order to be quantitative, we can calculate the growth rate of the
different modes, both analytically and numerically. If the initial spherical
symmetry is perturbed by a small amount, some instabilities will start
growing. We will study which instability modes are
going to prevail during the front evolution.

% \begin{figure}[htbp]
% \begin{center}
% \includegraphics[scale=0.3,angle=180]{parm_10_10_25}
% \end{center}
% \caption{Shape of the plasma at $t=0.0142$ for $Q=-25$, $E=0$ and
%$\epsilon=0.05$. The initial spherical symmetry is perturbed with the spherical
%harmonic $Y_{10,10}$}
% \label{arm_10_10}
% \end{figure}

We consider now a spherically expanding plasma, representing a corona discharge, with ${Q(t)}<0$ so that
$E_{0}=\mbox{E}_{\nu }^{+}=\frac{Q(t)}{4\pi R(t)^2}<0$. Then, $R(t)$ is given as the solution of
\begin{equation}
\frac{dR}{dt}=-\left( \frac{Q(t)}{4\pi R}+2\varepsilon \right) \frac{1}{R}%
+2\varepsilon ^{\frac{1}{2}}\sqrt{\alpha (|\mbox{E}_{\nu }^{+}|)}.
\label{r0d3}
\end{equation}
If $Q(t)=Q$ it is easy to check that
\begin{equation}
R(t)\approx \left( \frac{3\left\vert Q\right\vert }{4\pi }t\right) ^{\frac{1%
}{3}},
\end{equation}%
for the early stages of the discharge and as long as $R\quad \ll \frac{|Q|}{\varepsilon }$. This is in agreement
with predictions based on continuum streamer models \cite{aft1},\cite{r3}.
If the position of the front as well as the charge density are changed by a
small amount, the perturbed quantities can be parameterized as
\begin{eqnarray}
r(\theta ,\phi ,t) &=&R(t)+\delta S(\theta ,\phi ,t), \label{pert2d3} \\
\sigma (\theta ,\phi ,t) &=&-\frac{Q(t)}{4\pi R^{2}(\theta ,\phi ,t)}+\delta
\Sigma (\theta ,\phi ,t),
\end{eqnarray}%
where $\delta $ is a small
parameter. The angles $\theta $, and $\phi $ are the usual spherical
coordinates. For convenience we write the surface perturbation in terms of
spherical harmonics as
\begin{equation}
S=\sum_{l=1}^{\infty }\sum_{m=-l}^{l}s_{lm}(t)Y_{lm}(\theta ,\phi ),
\label{snd3}
\end{equation}%
and the surface charge density perturbation as
\begin{equation}
\Sigma =-\sum_{l=1}^{\infty }\sum_{m=-l}^{l}\left( \frac{2l+1}{R}\,b_{lm}+%
\frac{Q(t)}{4\pi R^{2}}\frac{l+1}{R}\,s_{lm}\right) Y_{lm}(\theta ,\phi )
\label{sigmand3}
\end{equation}%
where the coefficients $s_{lm}(t)$ and $b_{lm}$ have to be determined.
Making a standard expansion of the dynamics contour model equations %
\eqref{bb1} and \eqref{sigmaeq}, up to linear terms, we get the equations
for the particular mode evolution
\begin{equation}
\begin{split}
\label{eqSlm}
\frac{ds_{lm}}{dt}& =\Biggl[\varepsilon ^{\frac{1}{2}}\frac{\sqrt{\alpha _{0}%
}\,\mathrm{sign}(Q(t))}{|E_{0}|}\biggl(1+\frac{1}{|E_{0}|}\biggr)-1\Biggr]%
\frac{(l+1)}{R}\,b_{lm} \\
& +\Biggl[\varepsilon ^{\frac{1}{2}}\sqrt{\alpha _{0}}\left( 1+\frac{1}{%
|E_{0}|}\right) -E_{0}-\frac{\varepsilon (l+2)}{R}\Biggr]\frac{(l-1)}{R}%
\,s_{lm},
\end{split}%
\end{equation}%
\begin{equation}
\begin{split}
\label{eqblm}
& \frac{db_{lm}}{dt}=\frac{(l^{2}-1)E_{0}}{(2l+1)R}\left[2E_{0}+%
\frac{(l+4)\varepsilon }{R}-(\varepsilon \alpha _{0})^{\frac{1}{2}}\,\biggl(3+%
\frac{1}{|E_{0}|}\biggr)\right] s_{lm} \\
& -\frac{I(t)(l+1)}{4\pi R^{2}(2l+1)}s_{lm}+\Big[\frac{(l^{2}+4l+2)}{(2l+1)}%
E_{0}+\frac{2\varepsilon }{R} \\
& -\varepsilon ^{\frac{1}{2}}\frac{\sqrt{\alpha _{0}}}{(2l+1)}\biggl(%
l^{2}+6l+3+\frac{(l+1)^{2}}{|E_{0}|}\biggr)-\frac{lR}{(2l+1)\varrho }\Big]%
\frac{b_{lm}}{R}.
\end{split}%
\end{equation}

\begin{figure}[htbp]
\begin{center}
\includegraphics[scale=0.43]{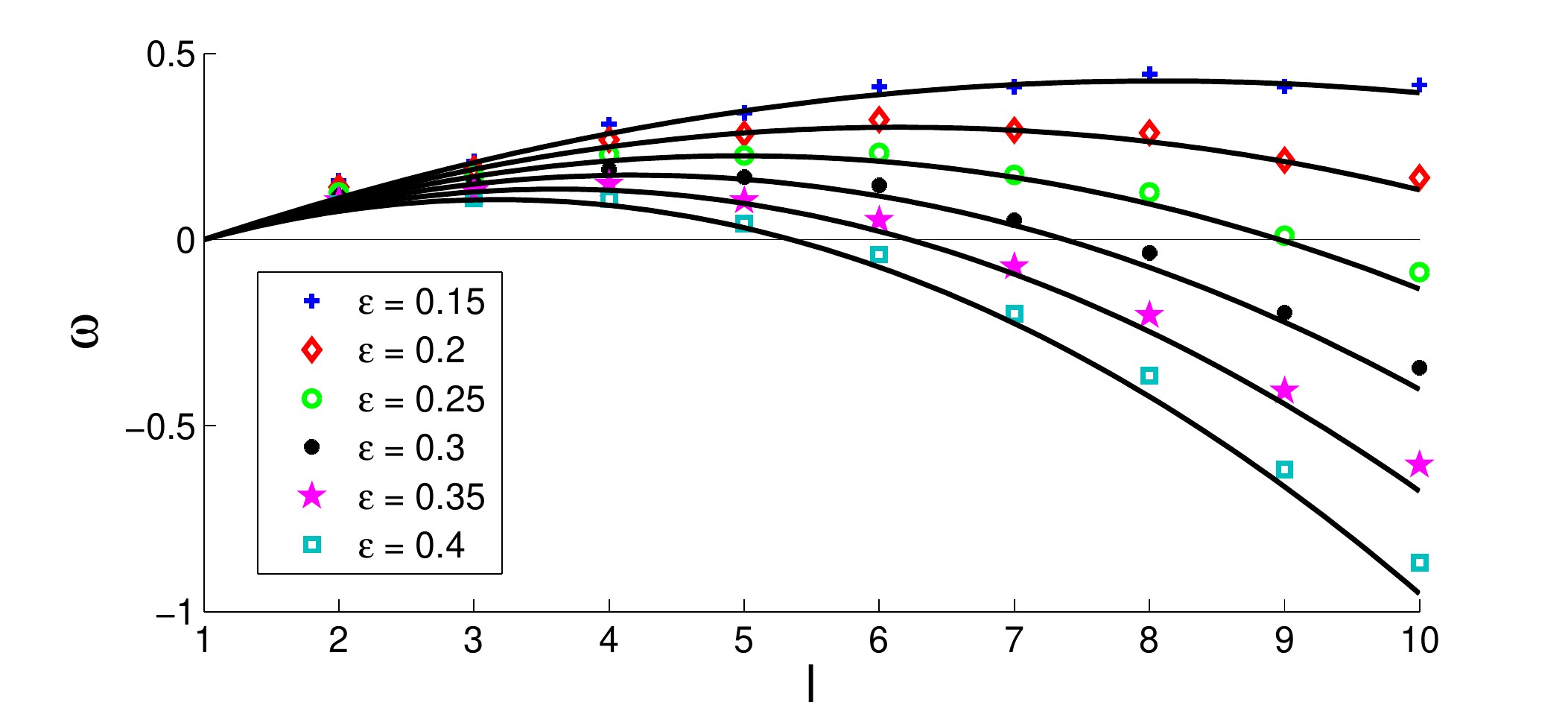}
\end{center}
\caption{Analytical (in black) and numerical dispersion curves for different
values of the diffusion coefficient $\varepsilon$. The abscissa
corresponds to the spherical harmonics number, and the ordinate to the
growth rate for that mode.}
\label{disp}
\end{figure}

We can get information about the growth of different modes by analyzing two
special limits. First we study the limit of ideal conductivity. It
corresponds to $\varrho \rightarrow 0$, and hence, from \eqref{eqblm}, we
can conclude that $b_{lm}\rightarrow 0$. This is the case when in the limit
of very high conductivity, the electric field inside goes to zero ($E_{\nu
}^{-}\rightarrow 0$), as we approach to the behavior of a perfect conductor.
If we consider that $Q(t)=Q_{0}$ is constant or its variation in time is
small compared with the evolution of the modes (which also implies $%
I(t)\rightarrow 0$), and the same for the radius of the front $R(t)=r_{0}$,
we look for a solution $s_{lm}=\exp (\omega t),\,\varphi _{n}=0$, to %
\eqref{eqSlm}, and get a discrete dispersion relation of the form
\begin{equation}
\omega =\Biggl[\varepsilon ^{\frac{1}{2}}\sqrt{\alpha _{0}}\left( 1+\frac{1}{%
|E_{0}|}\right) -E_{0}-\frac{\varepsilon (l+2)}{r_{0}}\Biggr]\frac{(l-1)}{%
r_{0}}, \label{disp1}
\end{equation}%
with a maximum at
\begin{equation}
l=l_{max} \simeq\frac{|E_{0}|r_{0}}{2\varepsilon}, \label{lmax}
\end{equation}%
for $\varepsilon\ll 1$. For a small enough conductivity, $\varrho \rightarrow \infty $, we find $%
b_{lm}=-E_{0}s_{lm}(l+1)/(2l+1)$, and with $s_{lm}=\exp (\omega t)$, %
\eqref{eqSlm} yields
\begin{equation}
%\begin{equation*}
\omega =\Biggl[\varepsilon ^{\frac{1}{2}}\sqrt{\alpha }_{0}\,\biggl(1+\frac{1%
}{|E_{0}|}\biggr)-E_{0}\Biggr]\frac{(l^{2}-3l-2)}{(2l+1)r_{0}}-\frac{%
\varepsilon (l+2)(l-1)}{r_{0}^{2}},
%\end{equation*}
\label{disp2}
\end{equation}%
with a maximum at
\begin{equation}
l=l_{max} \simeq\frac{|E_{0}|r_{0}}{\varepsilon}, \label{lmax2}
\end{equation}%
for $\varepsilon\ll 1$. Note that the dispersion relation does not depend on $m$. The finite
resistivity cases lay between those limits. In fig \ref{disp} we have plotted
the analytical curves given by \eqref{disp1} for different values of
$\varepsilon$ and the
results of numerical calculations for a perfect conductor.

The dispersion curve allows to predict the expected number of branches that
will develop. Each branch will undergo also a further split and so on
propagating to the smaller scales. However, it cannot run forever, as there
is a limitation and the model does not take into account the energy
radiated, the heat exchange, and the other phenomena that will start to play
an important role at later stages of the discharge.

%The discussion above concerns negative streamers, where electrons are accelerated as they move towards the front, produce new electrons by impact ionization and finally accumulate at the front. In positive streamers, electrons are removed from the front and the net charge is positive. Since electrons are accelerated in the opposite direction to the front, impact ionization contribute to decrease the net front charge and its velocity. The interface evolution equation is, instead of \eqref{bb1},
% \begin{equation}
%v_{N}=\mbox{E}_{\nu }^{+}+2\sqrt{\varepsilon\alpha(|\mbox{E}_{\nu }^{+}|)}%
%-\varepsilon'\kappa,  \label{bb1p}
%\end{equation}
%where $\varepsilon'=D_{ion}/D_0$. Since positive ions diffuse much less than negative electrons,
%then $\varepsilon'\ll\varepsilon$. This has an immediate consequence in the dispersion
%relations for spherically expanding streamers and, consequently, in the mode of maximum growth which is then given by
%\begin{equation}
%l=l_{max}^{+} \simeq\frac{|E_{0}|r_{0}}{2\varepsilon'}, \label{lmaxp}
%\end{equation}%
%instead of \eqref{lmax}. Since $l_{max}^{-}\ll\l_{max}^{+}$, we conclude that instabilities
%develop with much higher modes in positive streamers and hence filaments should be
%much thinner and dense in this case. Moreover, since $\omega$ grows quadratically with $\l$ for large values of $l$, we find much faster growth of filaments in positive streamers than in negative streamers and hence a faster speed of
%propagation. This is exactly what is observed in experiments \cite{Raether,Briels}.

The results presented in this work confirm the hypothesis that at the earlier
stages of an electric discharge, the main driving forces are diffusion and
electrical drift, first anticipated in \cite{ME}. The expressions obtained
for the growth rate of the modes, given by \eqref{disp1} and \eqref{disp2}
enables one to predict the number of forks that one can expect in an
electric discharge provided the electric field and the diffusion constant is
known by other means. But, the opposite can be worked out: from the numbers
of fingers observed, we can for example infer the electric field if the
charge density at the interface and the diffusion constant are known. This
has been done for the 2-D case \cite{Arrayas11}. For the 3-D case, the density
can be obtained from Stark's effect measurements, and effective diffusion
coefficient may be approximately calculated. These results contribute to
achieve one of the main goals, both in the laboratory and in nature, of the
current research in the area of electric discharges: bringing the field from a
qualitative and descriptive era to a quantitative one.

\textbf{Acknowledgments.} This work has been supported by the Spanish
Ministerio de Ciencia e Innovaci\'on under projects AYA2009-14027-C07-04, AYA2011-29336-C05-03 and MTM2011-26016.

% \begin{figure}[tbp]
% \centering
% \includegraphics[width=0.3\textwidth]{figure2} \includegraphics[width=0.3%
% \textwidth]{figure1} \label{descaga}
% \caption{Discharge}
% \end{figure}

\end{document}